\documentclass[twocolumn,aps,prb,longbibliography,superscriptaddress,floatfix]{revtex4-2}

\usepackage[T1]{fontenc}
\usepackage[utf8]{inputenc}
\usepackage[english]{babel}
\usepackage{color}
\usepackage{float}
\usepackage{braket}
\usepackage{amsmath}
\usepackage{bbding}
\usepackage{amstext}
\usepackage{amssymb}
\usepackage{graphicx}
\usepackage{bbold}
\usepackage{xcolor}
\usepackage{bm}
\usepackage[unicode=true,pdfusetitle,
bookmarks=true,bookmarksnumbered=false,bookmarksopen=false,
breaklinks=false,pdfborder={0 0 1},backref=false,colorlinks=false]
{hyperref}

\hypersetup{
	colorlinks,linkcolor=blue,citecolor=blue,urlcolor=blue}
\usepackage{soul}
\usepackage[normalem]{ulem}

\makeatletter
\date{}

\makeatother
\setul{0.5ex}{0.3ex}
\definecolor{Blue}{rgb}{0,0.0,1}
\setulcolor{Blue}
\usepackage{babel}

\begin{document} 

\title{Orbitronics in Two-dimensional Materials}
\author{Tarik P. Cysne}
\email{tarik.cysne@gmail.com}
\affiliation{Instituto de F\'\i sica, Universidade Federal Fluminense, 24210-346 Niter\'oi RJ, Brazil} 
\author{Luis M. Canonico}
\affiliation{Catalan Institute of Nanoscience and Nanotechnology (ICN2), CSIC and BIST, Campus UAB, Bellaterra, 08193 Barcelona, Spain} 
\author{Marcio Costa}
\affiliation{Instituto de F\'\i sica, Universidade Federal Fluminense, 24210-346 Niter\'oi RJ, Brazil} 
\author{R. B. Muniz}
\affiliation{Instituto de F\'\i sica, Universidade Federal Fluminense, 24210-346 Niter\'oi RJ, Brazil} 
\author{Tatiana G. Rappoport}
\email{tatiana.rappoport@inl.int}
\affiliation{Centro Brasileiro de Pesquisas Físicas (CBPF), Rua Dr Xavier Sigaud 150, Urca, 22290-180, Rio de Janeiro-RJ, Brazil}
\affiliation{Physics Center of Minho and Porto Universities (CF-UM-UP),Campus of Gualtar, 4710-057, Braga, Portugal}
\affiliation{International Iberian Nanotechnology Laboratory (INL), Av. Mestre José Veiga, 4715-330 Braga, Portugal}

\begin{abstract}
\section*{Abstract} Orbitronics explores the control and manipulation of electronic orbital angular momentum in solid-state systems, opening new pathways for information processing and storage. One significant advantage of orbitronics over spintronics is that it does not rely on spin-orbit coupling, thereby broadening the range of non-magnetic materials that can be utilized for these applications. It also introduces new topological features related to electronic orbital angular momentum, and clarifies some long-standing challenges in understanding experiments that rely on the conventional concept of valley transport. This review highlights recent advances in orbitronics, particularly in relation to two-dimensional materials. We examine the fundamental principles underlying the generation, transport, and dynamics of orbital angular momentum to illustrate how the unique properties of two-dimensional materials can promote orbitronic phenomena. We also outline potential future research directions and address some outstanding questions in this field.
\end{abstract}

\maketitle
\section{Introduction}

The control of current flow by manipulating the spin degrees of freedom has led to the emergence of spintronics. Significant advances in this field over the past few decades \cite{Fert2008, Sinova2015, Ahn2020, Avsar2020} have provided innovative solutions to enhance the diversity, functionality, and efficiency of devices used for information processing and storage. 

{Spin current is a key concept in the field of spintronics. The generation of pure spin current in time-reversal symmetric systems requires the presence of spin-orbit coupling (SOC).} This relativistic interaction, given by $\lambda\hat{{\bf S}}\cdot\hat{{\bf L}}$, couples the electronic spin operator $\hat{{\bf S}}$ with the orbital angular momentum (OAM) operator $\hat{{\bf L}}$. Its intensity $\lambda$ is material dependent and increases rapidly with the atomic number $Z$ of the constituent elements. 
A prominent spintronic phenomenon is the spin Hall effect (SHE), in which a transverse spin current is generated in response to a longitudinal applied electric field \cite{Dyakonov1971, Hirsch1999, Sinova2004}. In addition to producing pure spin currents for practical applications in spintronic devices, the SHE may serve as a reliable identifier of the quantum spin Hall insulating phase. This topological phase of matter has spurred a considerable amount of research activity in condensed matter physics over the past two decades \cite{Bernevig2006, Kane2005, Konig2007}. However, the need for a strong SOC may greatly limit the options for material selection. For instance, the relatively low atomic number of carbon presents a challenge to manipulating the electronic spin in pristine graphene. To address this limitation, one possible approach involves doping graphene with impurities that exhibit high spin-orbit coupling (SOC) values, or placing it sufficiently close to materials that can induce SOC in graphene through the proximity effect, in order to improve its potential application in spintronics \cite{Han2014, Avsar2020}. Similar limitations occur in other materials composed of elements with low atomic numbers \cite{Jansen2012}. However, materials with naturally high SOC are relatively rare, and their extraction processes often incur significant environmental and economic costs \cite{Rappoport2023}.

It is worth noting that an electric field does not interact directly with the electronic spin but couples to the charge carriers and may affect their orbital angular momenta. This perturbation can propagate through the system, generating an OAM current that does not require the presence of SOC, and is not necessarily accompanied by a charge current. 
The use of electronic orbital angular momentum degrees of freedom for information processing and storage has given rise to the field of orbitronics, which has been evolving very rapidly in recent years. 

Orbitronics shares similarities with spintronics. Several key spintronic effects, including the SHE, the inverse spin Hall effect (ISHE), the spin Rashba-Edelstein effect (SREE), and spin pumping, among others, have corresponding orbitronic analogs \cite{Go2021a, Jo2024, Wang2024}. The ability to generate orbital angular momentum currents without relying on spin-orbit interaction significantly broadens the range of materials that can be employed in orbitronic applications.
For example, the orbital Hall effect (OHE), which describes the emergence of a transverse orbital angular momentum current in response to a longitudinally applied electric field, was originally predicted to occur in $p$-doped silicon, a material characterized by weak spin-orbit coupling (SOC) \cite{Bernevig2005}. For several years, few theoretical works explored the OHE foreseen by Bernevig, Hughes, and Zhang. Most of them focused on three-dimensional metallic materials \cite{Kontani2009, Tanaka2008, Tanaka2010, Zhang2005, Kontani2008}, with a single exception that examined OHE in graphane \cite{Tokatly2010}.
Initially, some of the challenges in the field of orbitronics stemmed from experimental difficulties in unambiguously identifying the accumulation and transport of OAM. On the theoretical side, the definition of OAM in periodic systems also presented significant hurdles \cite{Thonhauser2005, Xiao2005, Bianco2013}. For sometime, there was also a certain skepticism regarding the importance of orbital effects compared to spin effects. In part, because of a prevalent belief that OAM quenching would render orbital physics less significant. 
However, this perception has changed significantly in recent years due to theoretical and experimental advances in the field of orbitronics \cite{Park2011, Go2018, Sunko2017, Park2012}. In fact, it took almost two decades, since its theoretical prediction, for the first direct observation of OHE to be announced in titanium \cite{Choi2023}. Additional studies on other low-SOC materials followed \cite{Lyalin2023, Sala2023} and, more recently, OHE has also been reported in silicon \cite{Matsumoto2025}, which was the material proposed in 2005 to exhibit this effect \cite{Bernevig2005}.

The OHE enables the generation of orbital angular momentum currents in non-magnetic systems, paving the way for the development of orbitronic devices. Conversely, inverse OHE can be utilized to detect these currents in certain materials. Other non-equilibrium orbitronic effects play a crucial role in manipulating electronic OAM. Notable examples include orbital torque (OT) effects \cite{Go2020a, Go2020b, Bose2023, Fukunaga2023, Santos2023, Lyalin2024} and orbital Rashba Edelstein effect (OREE) \cite{Yoda2018, Johansson2024, Nikolaev2024, Chirolli2022}, also called orbital magnetoelectric effect. Along with their corresponding inverse phenomena linked to the Onsager reciprocity relations, they have introduced novel methods for manipulating orbital and spin angular momenta in nanostructures.

Although much of the research in orbitronics has focused on metallic three-dimensional systems, there is increasing interest in two-dimensional (2D) materials. The reduction in dimensionality significantly alters the crystalline field experienced by electrons in 2D materials compared to the bulk. This change affects orbital hybridizations, which can have a profound impact on orbital and electronic properties. For example, thin films of transition metal dichalcogenides (TMDs), with stoichiometry MX$_2$, where M represents a transition metal atom and X denotes a chalcogen atom, exhibit electronic properties that depend on their composition, crystalline stacking, and thickness.
Interestingly, these materials exhibit sizable orbital Hall conductivity plateaus within their semiconducting gaps, where the spin Hall conductivity vanishes \cite{Canonico2020a, Canonico2020b, Costa2023}. The H structural phase of MoS$_2$ is actually an orbital Hall insulator, characterized by an orbital Chern number $C_{L_z}$, which takes the value of one for the monolayer and two for the bilayer \cite{Cysne2021a, Cysne2022}. Furthermore, TMD monolayers in their 2H and 1T structural phases have also been classified as higher-order topological insulators (HOTI) and they all display a plateau in orbital Hall conductivity within the band gap \cite{Zeng2021, Qian2022, Costa2023}.

2D materials represent an exciting platform for research in orbitronics, particularly those of the van der Waals type. They can be exfoliated to produce ultrathin films down to monolayer thickness, which can be easily placed onto various substrates to create hetrostructures with unique orbitronic features \cite{He2020, Bhowal2020, Cysne2023, Son2019, Seyler2018, Zhong2017}.

This review highlights significant advances in the orbitronics of 2D materials.
Section ``Theoretical background'', provides an overview of some basic concepts underlying orbitronic effects. 
In Section ``OAM in 2D Materials'', we discuss observations of orbital textures and some theoretical formulations of the orbital angular momentum operator, emphasizing their use in two-dimensional materials. 
Section ``OHE in 2D Materials'', brings attention to progress related to the orbital Hall effect in two-dimensional materials. In Section ``Controlling OAM'',  we provide an overview of the current landscape of the orbital Rashba-Edelstein effect (OREE), and orbital torques (OT), as explored by researchers in the field of two-dimensional materials.
Section ``Effects of Disorder'', explores the consequences of disturbances caused by impurities and defects, with a particular emphasis on orbital relaxation processes and extrinsic contributions to the OHE and OREE.
In the concluding section, we provide our insights into potential future research directions for orbitronics in 2D materials.

\section{Theoretical background}

\subsection*{Orbital Hall effect}
The first proposal for the orbital Hall effect was presented by Bernevig \textit{et al.}\cite{Bernevig2005}. In their seminal work, they introduced it as an effect analogous to the spin Hall effect, but centered on the transport of OAM. Later, Go \textit{et al.} in Ref. \cite{Go2018} proposed a microscopic mechanism that explains how OAM can be transported even in non-magnetic and centrosymmetric systems. In these systems, while \emph{equilibrium} OAM is quenched, an external electric field induces a \emph{non-equilibrium} { OAM} texture, leading to a transverse orbital current. Within linear response theory, the OHE can be mathematically expressed as:
\begin{eqnarray}
\mathcal{J}_y^{L_z}=\sigma^{L_z}_{yx}\mathcal{E}_x, \label{OHE}
\end{eqnarray}
where $\mathcal{J}_y^{L_z}$ represents the current density component along the $\hat{y}$ direction of the OAM component $L_z$, induced by an electric field with intensity $\mathcal{E}_x$ applied in the $\hat{x}$ direction. 

\begin{figure}[t]
    \centering
    \includegraphics[width=0.98\linewidth,clip]{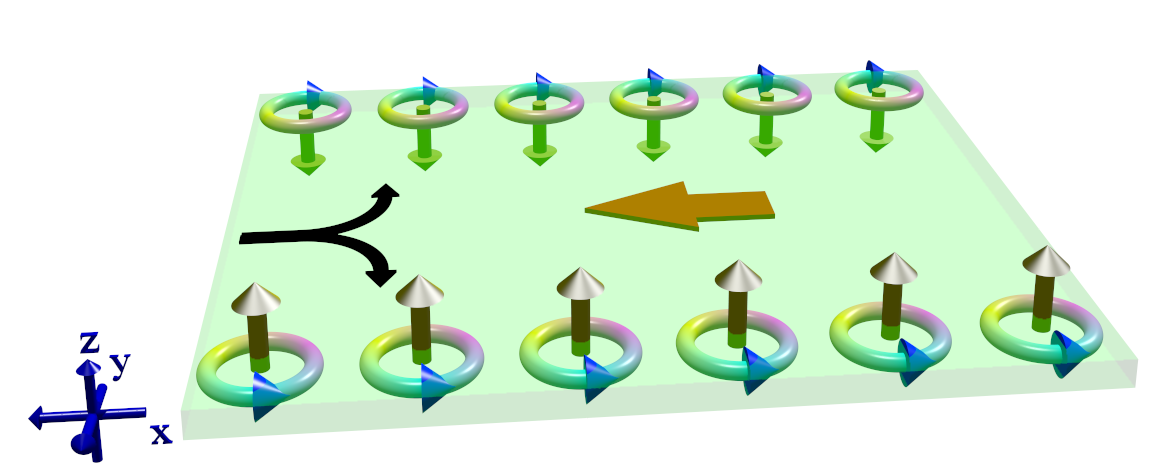}
        \caption{\textbf{OHE in a 2D system.} {Schematic representation of the orbital Hall effect [Eq.(\ref{OHE})]. A longitudinal charge current (black arrow) driven by an external electric field (red arrow) leads to a transverse flow of orbital angular momentum (OAM). This orbital current results in the accumulation of opposite OAM at the lateral edges of the sample, represented by golden arrows and phase rings. The color gradient in the rings indicates the phase structure of the electron wavefunctions associated with nonzero OAM. The splitting of the charge current line is purely pictorial, used to illustrate the opposite deflection of electrons carrying different signs of OAM. No transverse charge current is present.}}
    \label{fig:OHE}
\end{figure}

Fig. \ref{fig:OHE} shows a schematic representation of the OHE and the resulting accumulation of OAM at the edges of a 2D system due to the transverse orbital current. {It is worth mentioning that the orbital current is a quantum mechanical expectation value of the OAM flux carried by the conduction electrons; it does not represent a transverse flow of charge carriers. Rather, it describes how electrons, while moving longitudinally under an electric field, acquire OAM, resulting in a net angular momentum flow without any accompanying net charge transport in the transverse direction.}

This effect can occur in the absence of SOC as well as in the presence of spatial inversion and time-reversal symmetry. Within the picture of the mechanism introduced in Ref. \cite{Go2018}, SOC partially transfers the transverse orbital current to the spin sector [$\sigma^{L_z}_{yx}\rightarrow\sigma^{S_z}_{yx}$], giving rise to SHE as a secondary phenomenon resulting from the more fundamental OHE. Other theoretical aspects of the OHE are detailed in recent reviews \cite{Go2021a, Jo2024, Wang2024}. 

One may generalize equation \ref{OHE} to include other matrix elements of the orbital conductivity tensor $\sigma^{L_{\delta}}_{\mu,\nu}$. This tensor obeys the same symmetry properties as the spin conductivity tensor \cite{Roy2022, Seemann2015}. However, the physical picture elaborated in Ref. \cite{Go2018} applies to the elements of Eq. (\ref{OHE}) and its cyclic permutations in the Cartesian indices $x, y$ and $z$.

The current density operator is defined by,
\begin{eqnarray}
\hat{{\bf J}}^{L_z}=\frac{1}{2}\left(\hat{{\bf v}} \hat{L}_z+\hat{L}_z \hat{{\bf v}}\right), \label{JLz}
\end{eqnarray}
where $\hat{{\bf v}}$ and $\hat{L}_z$ represent the electronic velocity and z-component of the OAM operators, respectively. 

\begin{figure}[h]
    \centering
    \includegraphics[width=0.99\linewidth,clip]{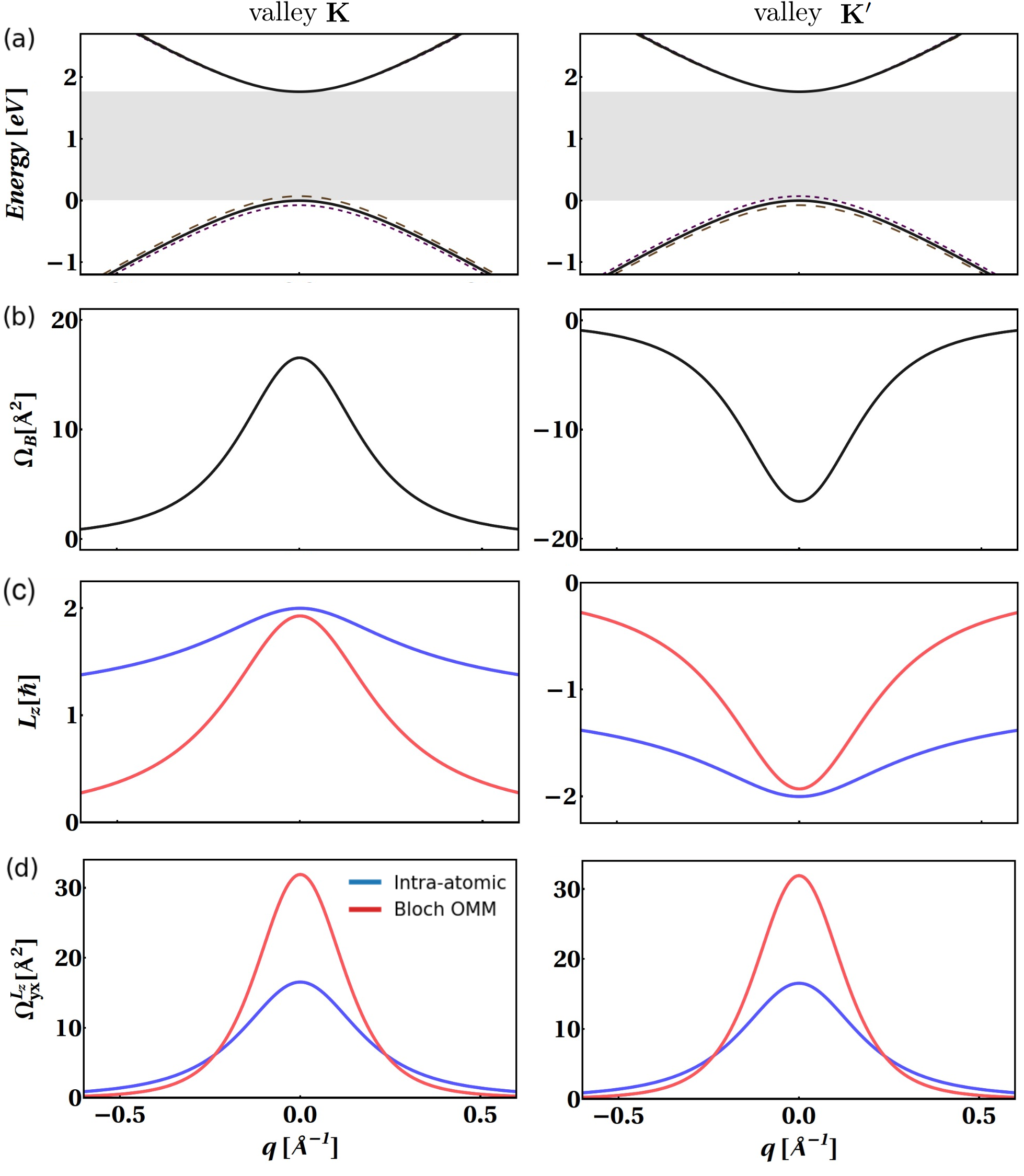}
        \caption{\textbf{Berry and Orbital Berry Curvature in 2H-TMD Monolayers.} {(a) Schemmatic illustration of the calculated energy bands for non-centrosymmetric 2H-TMD monolayer (a typical massive Dirac material) around the bottom of its conduction band and the top of the valence bands located at ${\bf K}$ and ${\bf K'}$ valleys of Brillouin zone (BZ). The black curve shows the spectrum without considering spin-orbit coupling (SOC). The dashed brown ($s=\uparrow$) and dotted purple ($s=\downarrow$) curves show the spin splitting of the valence bands caused by SOC. (b) The valence band Berry curvature near the valleys of the BZ. (c) The z-component of orbital angular momentum (OAM) near the valleys of the BZ. (d) The valence band orbital Berry curvature near the valleys of the BZ for a 2H-TMD monolayer. In the panels (c) and (d), the blue curves represent the results in the intra-atomic approximation, while the red curves show the results for OAM originating from the magnetic moment of Bloch states \cite{Cysne2022}. In panels (b)–(d), the SOC of the TMD is not considered, and a spin-degeneracy factor is included in the results for the Berry and orbital Berry curvatures. The low-energy theory of 2H-TMD monolayer with parameters for MoS$_2$ is used in these results.}}
    \label{fig:OBC}
\end{figure}

For 2D systems, the orbital Hall conductivity (OHC) $\sigma^{L_z}_{yx}$ can be expressed as
\begin{eqnarray}
\sigma^{L_z}_{yx} = e\sum_{n}\int_{\rm BZ}\frac{d{\bf k}}{(2\pi)^2}f(\epsilon_{n{\bf k}})\Omega^{L_z}_{yx,n}({\bf k}),\label{eqn:OHEBC}
\end{eqnarray}
where the orbital Berry curvature is 
\begin{eqnarray}
\Omega^{L_z}_{yx,n}({\bf k})=2\hbar\sum_{m\neq n}\text{Im}\left[\frac{\langle u_{n{\bf k}}\big|\hat{{\rm J}}^{L_z}_{y,{\bf k}}\big|u_{m{\bf k}}\rangle \langle u_{m{\bf k}}\big|\hat{{\rm v}}_x({\bf k})\big|u_{n{\bf k}}\rangle}{(\epsilon_{n{\bf k}}-\epsilon_{m{\bf k}}+i0^+)^2}\right]. \nonumber \\
\label{BerryCurvature}
\end{eqnarray}
Here, the components $\alpha=x,y$ of the velocity operators may be obtained by $\hat{{\rm v}}_{\alpha}({\bf k})=\hbar^{-1}\partial \hat{H}_{\bf k}/\partial {\rm k}_{\alpha}$, $f(\epsilon)$ denotes the usual Fermi-Dirac distribution function and $|u_{n{\bf k}}\rangle$ is the cell-periodic part of the Bloch states with eigenvalue $\epsilon_{n{\bf k}}$. {Figure~\ref{fig:OBC} illustrates the valence-band orbital Berry curvature for a monolayer of 2H-TMD, a prototypical multi-orbital Dirac material discussed in this review. In solids, the Berry curvature plays a role analogous to a magnetic field in momentum space, encoding how electronic wavefunctions twist or acquire geometric phase as the crystal momentum changes \cite{Xiao2010, Lesne2023}. The orbital Berry curvature refines this concept by taking into account the orbital character of the wavefunctions—effectively weighting the conventional Berry curvature by the expectation value of the OAM operator \cite{Go2018}. It captures how electrons with specific orbital identities respond to external fields, and directly contributes to orbital transport phenomena such as the OHE. While the conventional Berry curvature of monolayer 2H-TMD valence bands has opposite signs at the ${\bf K}$ and ${\bf K}'$ valleys—leading to effects like the valley Hall effect—the orbital Berry curvature, as shown in Fig. \ref{fig:OBC} (d), has the same sign at both valleys. This distinction is crucial, as it gives rise to a net orbital Hall conductivity even in centrosymmetric systems that preserve time-reversal symmetry \cite{Cysne2021a, Cysne2022}, and is responsible for the emergence of the orbital Hall insulating phase discussed in later sections \cite{Canonico2020a}.}

\subsection*{Orbital Rashba Edelstein effect}

\begin{figure}[t]
    \centering
    \includegraphics[width=0.98\linewidth,clip]{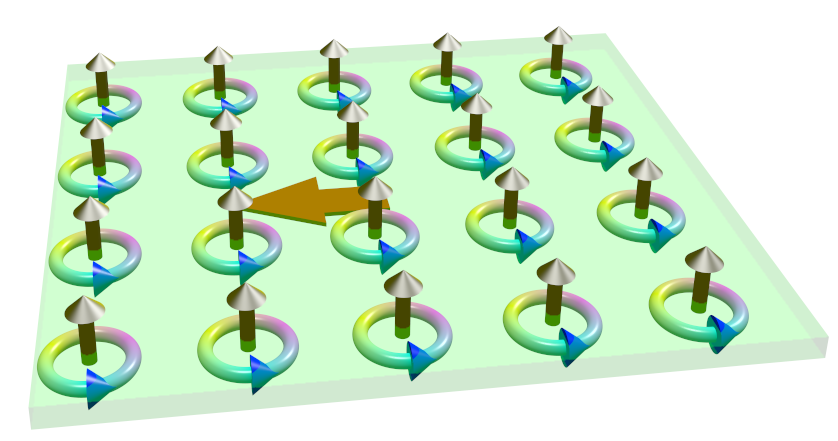}
        \caption{\textbf{OREE in a 2D system.} Schematic representation of orbital Rashba-Edelstein Effect [component $\beta_{zx}$ of Eq.(\ref{OEE2})].  An electric current passing through the material induces an orbital magnetization (golden arrows) oriented perpendicular to the current. The circulating electrons, driven by the phase gradient of their wavefunctions (color gradient in the rings), give rise to the orbital angular momentum.}
    \label{fig:OEE}
\end{figure}

The OREE refers to the generation of orbital magnetization induced by an electric current. It is a Fermi-surface phenomenon that can occur in non-magnetic systems but requires the breaking of spatial-inversion symmetry \cite{Hayami2018}. The OREE can be regarded as the OAM analog of the well-known Rashba-Edelstein effect (REE), which is characteristic of systems without inversion symmetry and with strong SOC. Sometimes, the OREE is also called the kinetic magnetoelectric effect. A detailed clarification of the different terminologies used can be found in Ref. \cite{Osumi2021}. Unlike the OHE, the OREE has been observed and studied for a long time \cite{Ivchenko1978, Vorobev1979} and was rediscovered by the orbitronics community. It can be described by the  linear response formula
\begin{eqnarray}
\mathcal{M}_{\mu}^L=\sum_{\nu}\alpha_{\mu\nu}\mathcal{E}_{\nu}, \label{OEE1}
\end{eqnarray}
where the non-equilibrium OAM density in $\mu$-direction ($\mathcal{M}_{\mu}^L$) generated by the electric field applied in the $\nu$ direction ($\mathcal{E}_{\nu}$) is proportional to a matrix-element of the tensor $\alpha_{\mu\nu}$, 
\begin{eqnarray}
&&\alpha_{\mu \nu}=-e\tau_p\frac{\mu_B}{\hbar}\sum_n\int_{\rm BZ}\frac{d{\bf k}}{(2\pi)^2}\left[\frac{\partial f}{\partial \epsilon}\right]_{\epsilon=\epsilon_{n{\bf k}}}\nonumber \\
&& \hspace{19mm}\cdot\bra{u_{n{\bf k}}}\hat{L}_{\mu}\ket{u_{n{\bf k}}}\bra{u_{n{\bf k}}}\hat{\rm v}_{\nu}({\bf k})\ket{u_{n{\bf k}}},
\label{METensor}
\end{eqnarray}
where $\mu_B$ denotes the Bohr magneton {and $\tau_p$ represents the momentum relaxation time}. Eq. (\ref{METensor}) is strongly constrained by crystal symmetry \cite{Furukawa2021}.  Since OREE is a Fermi surface effect, the previous equation can be expressed in terms of the charge current as follows:
\begin{eqnarray}
\mathcal{M}_{\mu}^L=\sum_\nu\beta_{\mu\nu}\mathcal{J}_{\nu}, \label{OEE2}
\end{eqnarray}
where $\mathcal{J}_{\nu}$ is the charge current flowing in the $\nu$-direction and $\beta_{\mu\nu}=\alpha_{\mu\nu}/\sigma_{\nu\nu}$, with $\sigma_{\nu\nu}$ being the longitudinal conductivity. Note that {since $\alpha_{\mu\nu} \propto \tau_p$, the tensor $\beta_{\mu\nu}$ is independent of $\tau_p$}. Fig. \ref{fig:OEE} illustrates the OREE associated with the $\beta_{zx}$ tensor component in Eq. \ref{OEE2}.

There are 18 noncentrosymmetric point groups, known as gyrotropic groups, which allow finite matrix elements in the current-induced orbital magnetization tensor \cite{Furukawa2021}: $C_1$, $C_2$, $C_3$, $C_4$, $C_6$, $C_{1{\rm v}}$, $C_{2{\rm v}}$, $C_{3{\rm v}}$, $C_{4{\rm v}}$, $C_{6{\rm v}}$, $D_{2{\rm d}}$, $S_4$, $D_2$, $D_3$, $D_4$, $D_6$, $T$, $O$. In addition, there are 3 noncentrosymmetric point groups, $C_{\rm 3h}$, $D_{\rm 3h}$, and $T_{\rm d}$, that exhibit the necessary spatial inversion symmetry breaking but do not allow for current-induced magnetization \cite{Furukawa2021}. The shape of the response tensor for each of the 18 gyrotropic point groups can be found in the literature \cite{He2020b}. Many subtleties of the microscopic mechanisms underlying the OREE, along with details on the various nomenclatures used in the literature, are discussed in a recent specialized review \cite{Johansson2024}. Here, we adopt a simple and intuitive explanation based on the Heisenberg equation of motion for the induced OAM in a solid subjected to an applied dynamical electric field \cite{Salemi2019}: $\frac{d\hat{{\bf L}}^{\rm Ind}}{dt}=-\frac{i}{\hbar}\left[\hat{{\bf L}}^{\rm Ind}, \hat{V}(t) \right]=\hat{{\bf r}} \times e\boldsymbol{\mathcal{E}}(t)$, where $\hat{V}(t)=-e\boldsymbol{\mathcal{E}}(t)\cdot\hat{{\bf r}}$ is a slowly switch-on electric potential. This general quantum mechanical equation indicates that the OREE arises from the torque applied to local electric dipoles within the solid: $\boldsymbol{\mathcal{M}}^{L}\propto {\bf P}\times \boldsymbol{\mathcal{E}}$ \cite{Cysne2023, Cysne2021b}. 

\subsection*{Orbital torques}
{ From a phenomenological point of view, the electric control of magnetization can be understood via the SOC, which acts together with the crystal field as a source of magnetic anisotropy -responsible for the orientation of the magnetization and orbital moments in ferromagnets \cite{Bruno1989,Coey2021}. Consequently, a variation in the SOC can trigger magnetization dynamics due to the instantaneous change in the system's magnetic anisotropy. Given by $\mathcal{H}_{SOC} = \lambda(\bm{r})\hat{{\bf L}}\cdot\hat{{\bf S}}$, the SOC depends on three essential parameters: the electrostatic potential at the vicinity of the nucleus $\lambda(\bm{r})$, the spin $\hat{{\bf S}}$, and the OAM of the electrons $\hat{{\bf L}}$. Thus, changes in the SOC are given by
\begin{align}
    \delta \mathcal{H}_{SOC} &= \delta\lambda(\bm{r})\hat{{\bf L}}\cdot\hat{{\bf S}} +\lambda(\bm{r})\hat{{\bf L}}\cdot\delta\hat{{\bf S}}\nonumber\\  {}& + \lambda(\bm{r})\delta\hat{{\bf L}}\cdot\hat{{\bf S}}\label{eq:SOCEnergy},
\end{align}
\noindent where the first term in Eq. \eqref{eq:SOCEnergy} encapsulates the variation of the atomic Coulomb potential due to the motion of the ions, the second term is related to changes in the spin density of the system, and the third term capture the changes in the OAM density. The first mechanism has been recently explored by Han \textit{et al.} \cite{Han2023a} as a way of using the lattice to harvest spin and orbital moments. The second channel connects directly to the spin-orbit torques \cite{Manchon2019}, whereas the third describes the orbital torque in which a non-equilibrium orbital density triggers the magnetization dynamics. This mechanism was proposed by Go and Lee in Ref. \cite{Go2020a}. The system they used was a bilayer comprising a non-magnetic material (NM) and a ferromagnetic material (FM) with strong SOC. The NM material is formed by $s$-$p$ hybridized orbitals (similar to the model used in Ref. \cite{Go2018}), while the FM includes $d$ orbitals. The equilibrium magnetization ${\bf M}$ of FM interacts with electronic spins density via exchange coupling $H^{\rm FM}_{\rm xc}=\left(J/\hbar\right){\bf M}\cdot \hat{{\bf S}}$. The SOC of FM is given by $H_{\rm so}^{\rm FM}=\left(\alpha^{\rm FM}_{\rm so}/\hbar^2\right)\hat{{\bf L}}\cdot\hat{{\bf S}}$. When an electric field is applied, a transverse orbital current flows in the NM material due to the OHE. This orbital current is then transferred to the FM, injecting an OAM density, and the angular momentum of this nonequilibrium distribution is transferred to the spins via SOC. This process induces a torque in the magnetization of the FM material. The torque is given by
\begin{eqnarray}
{\bf T}=\frac{J}{\hbar}{\bf M}\times \langle \hat{{\bf S}}\rangle^{\rm FM}, \label{OT}
\end{eqnarray}
where $\langle \hat{{\bf S}}\rangle^{\rm FM}$ is the spin accumulation produced in FM. Note that although the expression of Eq. (\ref{OT}) for OT depends on spin accumulation, it results from the injection of the OAM due to OHE in the NM material described by the third term in \eqref{eq:SOCEnergy}. This expression can be decomposed into field-like and damping-like components.  While the field-like component arises from the intra-band term of linear response theory, the damping-like term is associated with inter-band contribution \cite{Go2020a}. This phenomenon has been experimentally analyzed in various NM/FM bilayers and represents an important route for manipulating the OAM degree of freedom \cite{Lee2021, Hayashi2023, Yang2024}. The reciprocal effect, orbital pumping,  has been studied theoretically and its experimental evidence has also been reported \cite{Santos2023,Hayashi2024,Go2023,Han2023a, Abrao2025}. {It is worth mentioning that the model described above assumes that the magnetization of a ferromagnet is primarily attributed to the spin sector. In the cases where the ferromagnet has significant contributions from the orbital sector to its magnetization, the model can be generalized \cite{Aase2024}.}

\subsection*{Other orbital phenomena}
The effects described in Eqs.(\ref{OHE}-\ref{OT}), along with their reciprocal effects, can be viewed as building blocks for orbitronics devices within the generation-manipulation-detection paradigm, which is aimed at enabling efficient storage and processing of information using electronic OAM. Nevertheless, this has not exhausted the pool of orbitronic phenomena that could be useful for this task. Distinct mechanisms of orbital-charge conversion \cite{Santos2024}, {orbital Ruderman-Kittel-Kasuya-Yosida interaction \cite{Aase2024}}, and the interaction between OAM and pulsed light signals \cite{Seifert2023,Wang2023, Xu2024} are other topics currently under theoretical and experimental investigation. Here, we focused on the three building blocks, whose mechanisms and key references were outlined above.

\section{Orbital angular momentum in 2D materials}
\subsection*{Orbital textures}

The mechanism introduced in Ref. \cite{Go2018} to explain the OHE depends on the presence of orbital textures, $\bf{k}$-dependent orbital states, even in systems with inversion and time-reversal symmetry\cite{Han2023b}. These textures, which describe the spatial distribution of orbital character, are an important feature in many 2D materials. Within the framework of the intra-atomic approximation (see the next subsection), these textures are associated with $\bf{k}$-dependent hopping processes involving different orbitals. However, 2D materials with broken inversion symmetry can give rise to OAM textures, described by $\boldsymbol{\ell}_n({\bf k})=\bra{u_{n{\bf k}}}\hat{{\bf L}}\ket{u_{n{\bf k}}}$, where $\ket{u_{n{\bf k}}}$ represents the periodic part of the Bloch wave functions. Such OAM textures highlight the unique role of symmetry breaking in 2D materials.  They are found in various multi-orbital solids, including monolayers of 2H-TMDs \cite{Canonico2020a}, bismuthene \cite{Canonico2020b}, and borophene \cite{Lima2019}. 

Although the equilibrium OAM texture is quenched in non-magnetic and centrosymmetric systems, the presence of orbital textures can trigger a non-equilibrium OAM texture responsible for the occurrence of the OHE under the application of an electric field \cite{Go2018}. 
Experimental approaches based on analyzing angle-resolved photoemission spectroscopy signals have been developed to access the orbital and OAM textures of 2D materials and the surfaces of 3D materials \cite{Beaulieu2020,Mazzola2024}, applied to both equilibrium and non-equilibrium textures \cite{Beaulieu2020} and have been successfully used for in-depth studies of many compounds in the TMDs family \cite{Beaulieu2021, Schler2022}.
 
\subsection*{Multi-orbital v.s. single orbital materials}
In the low-energy regime, carbon $p_z$ orbitals are responsible for the electronic properties of graphene. This part of the Hilbert space, consisting of a single cubic harmonic state, does not support atomic OAM. Much of the discussion above is based on the intra-atomic approximation for the OAM operator, where the OAM of a solid is related to the OAM of individual atoms. This approach, also called the atom-centered approximation (ACA), becomes straightforward when the electronic Hamiltonian is expanded using a tight-binding basis, and it is commonly used in orbitronics. In graphene, the dominance of the $p_z$ orbital causes the intra-atomic OAM to vanish, making orbitronic effects impossible under this approximation. A similar situation occurs in $s$-character band structures. However, electronic wave functions in solids are not always localized around atoms, and contributions to OAM from inter-site electron movement can be significant. This aspect has been explored in magnetic materials, where such contributions affect the \emph{equilibrium} orbital magnetization. The \emph{modern theory of orbital magnetization} \cite{Xiao2005, Thonhauser2005} has improved theoretical predictions for many magnetic systems and has a substantial impact on more exotic materials, significantly altering results compared to the intra-atomic approximation \cite{Hanke2016}.

For isolated bands, Kohn demonstrated that Bloch electrons have an intrinsic orbital magnetic moment (OMM) \cite{Kohn1959}. Later, Chang and Niu reinterpreted this intrinsic OMM as the self-rotation of an electronic wave packet using a semi-classical approach \cite{Chang1996}. Culcer et al. extended this concept by deriving a non-abelian (matrix) form of the OMM operator for nearly degenerate bands \cite{Culcer2005}. Pezo and coworkers further generalized the semi-classical expression of the OMM, incorporating matrix elements from the full Hilbert space \cite{Pezo2022}:
\begin{eqnarray}
&&m_{n,m}^z({\bf k})=\nonumber \\
&&\hspace{5mm} -i\frac{e}{2\hbar}\bra{\vec{\nabla}_{\bf k}u_{n\bf k}}\boldsymbol{\times}\left[\hat{H}_{\bf k}-\mathbb{1}\left(\frac{\epsilon_{n{\bf k}}+\epsilon_{m{\bf k}}}{2}\right)\right]\ket{\vec{\nabla}_{\bf k}u_{m\bf k}}, \nonumber \\
\label{OMM}
\end{eqnarray}
where, $\vec{\nabla}_{\bf k}=\partial_{k_x}\vec{x}+\partial_{k_y}\vec{y}$, and $\ket{u_{n(m){\bf k}}}$ is the periodic part of Bloch wave-function for the band with energy $\epsilon_{n(m){\bf k}}$. Eq. (\ref{OMM}) is analogous to the expression obtained from the semi-classical approach but is not restricted to the nearly degenerate subspace. The same expression was obtained by Ref. \cite{Gobel2024}. As pointed out in Ref. \cite{Pozo2023}, an additional term must be added to Eq. (\ref{OMM}) to ensure gauge invariance in the case of {spaced} energy bands. Recently, an additional quantum term has been predicted \cite{Liu2025}. 

{ An OAM operator originating from this Bloch OMM can be defined \cite{Chang2008, Chang1996,Bhowal2021}:
\begin{eqnarray}
L^z_{n,m}({\bf k})=-\left(\frac{\hbar}{\mu_Bg_L}\right)m_{n,m}^z({\bf k}), \label{LzBloch}
\end{eqnarray}
where, $\mu_B$ is the Bohr magneton and $g_L=1$ is the Land\'{e} $g$-factor. This operator accounts for \emph{both} intrasite and intersite contributions to OAM, going beyond the intra-atomic contribution \cite{Souza2008}.}

{The formulation of OHE based on Bloch OMM \cite{Bhowal2021} has shown consistency with the intra-atomic approximation in predicting an orbital Hall insulating phase for bilayer of 2H-TMDs \cite{Cysne2022}. Nevertheless, the plateaus predicted by the intra-atomic approximation and the OMM approach have distinct heights and dependencies on the spectral energy band gap. Later, similar studies on other multi-orbital 2D materials were carried out \cite{Pezo2022}. Interestingly, even for single-orbital solids such as graphene \cite{Bhowal2021} and $s$-orbital lattices \cite{Busch2023}, the formulation based on the OMM approach gives rise to a finite OHE. In these situations, the intra-atomic approximation yields a vanishing OHE, and all contributions must be attributed to the intersite movement of electrons. A pictorial representation of this movement was described in Ref. \cite{Busch2023}. The OREE was also studied in 2D materials using the formulation of Bloch OMM \cite{shi2019, He2020,  Lee2024, Bhowal2020}. Studies on OT are often conducted using the intra-atomic approximation, which simplifies numerical calculations.} 

With the growing interest in comparing theoretical and experimental results, where disorder is an inherent feature of all materials, there is an increasing need for alternative formulations of the OAM operator in real space. One such approach is a basis-independent formulation using Green functions, expressed as:

\begin{eqnarray}
 &&\hat{L}_{\gamma} =  -\varepsilon_{\alpha\beta\gamma}\frac{ie\hbar^2}{4g_{L}\mu_B}\int dE\Big[\text{Re}(G^{+}(\hat{H},E))\hat{{\rm v}}_{\alpha}\delta(\hat{H}-E)\hat{{\rm v}}_{\beta}\nonumber\\
 && \hspace{25mm}+ \hat{{\rm v}}_{\alpha}\delta(\hat{H}-E)\hat{{\rm v}}_{\beta}\text{Re}(G^{-}(\hat{H},E))\Big],\label{eqn:RealSpaceOM}
\end{eqnarray}
that was introduced by Canonico \textit{et al.} \cite{Canonico2024}. Within this approach, the off-diagonal elements of the position operator $\bm{r}$ are computed from the perturbation theory methods as $\langle i|r_{\alpha}|j\rangle = i\hbar \frac{\langle i| \hat{{\rm v}}_{\alpha}|j\rangle}{E_j - E_i}$ ($\alpha=x,y,z$), where $|i\rangle$ and $E_i$ are the $i$-th energy eigenstate of Hamiltonian $\hat{H}$ with eigenvalue $E_i$.  $G^{-}(\hat{H},E)$ and 
$G^{+}(\hat{H},E)$ are the advanced and retarded Green's functions respectively and $\varepsilon_{\alpha\beta\gamma}$ represent the Levi-Civita symbol. The formulation in Eq. \eqref{eqn:RealSpaceOM} has been used to calculate the OHE due to the OMM in real space for disordered systems \cite{Canonico2024}. The expressions in Eqs. \eqref{OMM}, \eqref{LzBloch} and \eqref{eqn:RealSpaceOM} are used to describe the transport of OAM via the orbital current definition shown in Eq. \eqref{JLz}.

The expression of Eq. \ref{eqn:RealSpaceOM} is inspired 
 in the work of Bianco and Resta \cite{Bianco2013} that represents the macroscopic equilibrium orbital magnetization of a magnetic system as an integral of a local quantity and is equivalent to the more usual expression of the orbital magnetization, where the integration is performed in the reciprocal space:
 \begin{widetext}
\begin{eqnarray}
    &&M_z =  -\frac{e}{2\hbar c} \int \frac{d\bm{r}}{A}\text{Im} \bra{\bm{r}} \big| \hat{H}-\mu\mathbb{1}\big|\left[\bm{r}, \mathcal{P}\right]\boldsymbol{\times}\left[\bm{r}, \mathcal{P}\right]\ket{\bm{r}} \nonumber \\
    &&\hspace{7mm} =-\frac{e}{2\hbar c} {\rm Im} \sum_{n}\int_{\epsilon_{n{\bf k}}\le\mu}\frac{d{\bf k}}{(2\pi)^2} \bra{\vec{\nabla}_{\bf k}u_{n{\bf k}}}\boldsymbol{\times}\left(\hat{H}_{\bf k}+\mathbb{1}(\epsilon_{n{\bf k}}-2\mu)\right) \ket{\vec{\nabla}_{\bf k}u_{n{\bf k}}}.
    \label{ModernTheory}
\end{eqnarray}
\end{widetext}
The first line of Eq.  \eqref{ModernTheory} represents the macroscopic orbital magnetization as an integral of a local quantity \cite{Thonhauser2005, Bianco2013}. $A$ is the area of the sample, $\mu$ is the chemical potential, $\mathcal{P} = \sum_{\epsilon_n<\mu}|\phi_n\rangle\langle \phi_n|$ is the projector over the occupied states, and $|\hat{H}-\mu\mathbb{1}|=\big(\hat{H}-\mu\mathbb{1}\big)\big(\mathbb{1}-2\mathcal{P}\big)$ \cite{Andreoni2020}. This expression exploits the spectral representation of the orbital moment due to its independence on the characteristics of the energy states and is suitable for computing the orbital moment in disordered materials. The second line in Eq. (\ref{ModernTheory}) in turn, is {the} usual expression that exploits the periodicity of clean systems and represents it as a bulk property \cite{Xiao2005, Thonhauser2005}, i.e., an integral of a geometric quantity over the BZ. Note that in both equations, the orbital magnetization does not depend on the explicit representation of the multi-orbital nature of the band structure. It depends only on the eigenstates and eigenvalues of the Hamiltonian of magnetic material.

\section{Orbital Hall effect in 2D materials}

The reduced dimensionality and unique structural properties of 2D materials support the emergence of topological effects. Their planar structure also enables the control of the crystal field environment and manipulation of the symmetries governing the electronic properties. This has been extensively explored in spintronics research on 2D materials. For instance, reducing the symmetry group of a solid can lead to the emergence of novel components in the spin-conductivity tensor \cite{Roy2022}. In many 2D materials, this can be achieved and controlled through strain or proximity effects \cite{Safeer2019a, Safeer2019b, Camosi2022, Vila2021, Cysne2018}. Similar behavior is expected in the OAM conductivity tensor \cite{Costa2023}, making these materials a flexible platform for investigating OAM transport properties. As a result, numerous studies have explored orbital transport in 2D materials \cite{Phong2019, Zeer2022, Barbosa2024, Bhowal2020a, Bhowal2020b, Fonseca2023,Cysne2021a,Cysne2021b,Cysne2022,Canonico2020a,Canonico2020b,Canonico2024,Costa2023,Veneri2024,Pezo2023,Sanchez2024,Sun2024, Faridi2025,Bhowal2021, Ji2024,Ji2023, Mu2021, Cysne2024a,Cysne2024b,Cysne2023b, Busch2023,Liu2024b, Chen2024,Li2024}. In this section, we highlight some of the most significant findings in this area, focusing on two key aspects in the orbitronics of 2D materials: the interplay between topological properties and OAM-related effects, and the reformulation of valleytronic concepts in light of the OHE.

\subsection*{Orbital topology}

The quantum spin Hall insulator is a topological phase where the bulk of the system has an energy band gap indexed by a $\mathbb{Z}_2$ topological invariant, which remains robust against smooth changes to the Hamiltonian. When the Fermi energy lies within this gap, linear response theory predicts a finite spin Hall conductivity, a hallmark of its non-trivial topology. Furthermore, creating boundaries in such systems reveals spin-polarized metallic edge states that carry the spin-Hall current, protected by time-reversal symmetry. This concept, introduced by Kane and Mele, is foundational in the study of topological insulators \cite{Kane2005}.

While these properties are well-established in the spin sector, their analogs in the orbital sector exhibit pronounced differences. Refs. \cite{Canonico2020a, Canonico2020b} demonstrated that a finite OHC plateau can arise in certain insulating phases of 2D systems (see Fig. \ref{fig:TMD-top}b for the OHC of a MoS$_2$ bilayer). The $p_x$-$p_y$ model on a honeycomb lattice, used to describe group-V materials grown on SiC substrates \cite{reis2017, li2018}, serves as an example \cite{ Canonico2020b}. In this model, the $p_z$ orbital is excluded from the low-energy sector, leaving $p_x$ and $p_y$ orbitals to dominate electronic transport. Such orbital filtering provides an example of manipulating the OAM nature of a solid by altering the crystalline environment, a possibility widely explored in 2D materials.

Monolayers of TMDs with a 2H structural phase exhibit broken inversion symmetry, as can be seen in Fig. \ref{fig:TMD-top}a if one considers only one layer. This leads to unique OAM-polarized edge states that cross the band-gap region when shaped into a nanoribbon with zigzag edges \cite{Liu2014, Cysne2023}, shown in Fig. \ref{fig:TMD-top}c. These edge states are responsible for transporting the OHC \cite{Cysne2023}. Notably, monolayers of 2H-TMDs also display an OHC plateau within the insulating gap \cite{Canonico2020a}. To further understand the origin of the OHC plateau and to differentiate it from effects purely related to inversion symmetry breaking, the OHC was calculated in bilayers of TMDs, which maintain inversion symmetry (see \ref{fig:TMD-top}a) \cite{Cysne2021a} and the plateau OHE persists. However, unlike the quantum spin Hall insulator, the orbital Hall insulator cannot be indexed by a $\mathbb{Z}_2$ topological invariant. Instead, the orbital Chern number, introduced in Refs. \cite{Cysne2021a, Cysne2022}, serves as a topological invariant for the orbital Hall insulating phase of 2H-TMDs. Monolayers of 2H-TMDs are characterized by an orbital Chern number $C^{\rm 1l}_{L_z}=1$, while bilayers have $C^{\rm 2l}_{L_z}=2$. In the low-energy regime, where the Hamiltonian of 2H-TMDs can be written as Dirac-like Hamiltonian, the OHC plateau within the intra-atomic approximation is proportional to the orbital Chern number, $\sigma^{L_z}_{xy} = \left( \frac{e}{2\pi} \right) 2C_{L_z}$ \cite{Cysne2022}. The edge states in these systems, while related to the orbital Chern number, do not enjoy the same topological protection as those in the quantum spin Hall effect due to the absence of a $\mathbb{Z}_2$ invariant. Subsequent studies have further explored this topological invariant in other TMDs, showing their susceptibility to topological phase transitions \cite{Ji2024, Ji2023}.

In Refs. \cite{Zeng2021, Qian2022}, it was proposed that monolayers of 2H-TMDs exhibit a higher-order topological insulator (HOTI) phase, characterized by the presence of corner states with fractional charge when the layer is fabricated in a flake geometry that preserves threefold rotational symmetry. These corner states manifest as discrete energy levels within the bulk band gap of the energy spectrum, indicating their topologically protected nature and distinguishing them from bulk or edge states (see Figure \ref{fig:TMD-top}d). This higher-order topology is closely connected to the multi-orbital nature of 2H-TMDs. Subsequently, a study \cite{Costa2023} linked this HOTI phase to a finite orbital Hall effect  within the insulating gap in numerous TMD compounds that crystallize in two distinct structural phases: the non-centrosymmetric 2H phase and the centrosymmetric 1T phase. The connection between HOTI and OHE in these TMD phases was justified using symmetry arguments and supported by a study with an effective Bernevig-Hughes-Zhang (BHZ) toy model. {Table \ref{table_TMD-Hall} shows the transverse orbital conductivity of TMD insulating monolayers  crystallized in the 1T and 2H structural phases. The structure of each material was obtained from the C2DB database \cite{Haastrup2018}. The transverse spin conductivity vanishes within the energy bandgap for all compounds listed in the tables.} Subsequent works have extended the connections between topology and the multi-orbital nature of solids, particularly between HOTI and OHE \cite{Barbosa2024, Hu2024}, in materials like 2D ferroelectrics and superconductors  \cite{Arouca2024}, as well as in 2D ferromagnets \cite{Chen2024}. Additionally, well-established topological phenomena, such as the quantum Hall effect, are being reinterpreted with advances in understanding the orbital angular momentum (OAM) of electrons in solids \cite{Gobel2024b}.

\begin{table*}[t]
    \centering
    \begin{minipage}{0.5\textwidth}
        \centering
        \begin{tabular}{||c c || c c | c c | c c | c c | c ||} 
		\hline
		2H-TMD & &  $\sigma^{L_y}_{yx}$ & &  $\sigma^{L_z}_{yx}$ & & $|\sigma^{L}_{yx}|$ & & $|\sigma^{S}_{yx}|$ & & $E_g (\text{eV})$ \\ [0.5ex] 
		\hline
		\hline
		MoS$_2$ & & $0.00$ & & $2.65$ & & $2.65$ & & $0.00$ & & $1.60$\\
		\hline
		MoSe$_2$ & & $0.00$ & & $2.63$ & & $2.63$ & & $0.00$ & & $1.34$\\ [0.5ex] 
		\hline
		MoTe$_2$ & & $0.00$ & & $2.47$ & & $2.47$ & & $0.00$ & & $0.95$\\ [0.5ex] 
		\hline
		CrS$_2$ & & $0.00$ & & $1.89$ & & $1.89$ & & $0.00$ & & $0.90$\\ [0.5ex] 
		\hline
		CrSe$_2$ & & $0.00$ & & $1.92$ & & $1.92$ & & $0.00$ & & $0.71$\\ [0.5ex] 
		\hline
		CrTe$_2$ & & $0.00$ & & $1.91$ & & $1.91$ & & $0.00$ & & $0.47$\\ [0.5ex] 
		\hline
		WS$_2$  & & $0.00$ & & $1.43$ & & $1.43$ & & $0.00$ & & $1.56$\\ [0.5ex] 
		\hline
		WSe$_2$ & & $0.00$ & & $1.63$ & & $1.63$ & & $0.00$ & & $1.27$\\ [0.5ex] 
		\hline
		WTe$_2$ & & $0.00$ & & $1.46$ & & $1.46$ & & $0.00$ & & $0.77$\\ [0.5ex] 
		\hline		
		TiS$_2$ & & $0.00$ & & $0.78$ & & $0.78$ & & $0.00$ & & $0.72$\\ [0.5ex] 
		\hline
		TiSe$_2$ & & $0.00$ & & $0.75$ & & $0.75$ & & $0.00$ & & $0.53$\\ [0.5ex]
		\hline 
        \end{tabular}
    \end{minipage}%
    \begin{minipage}{0.5\textwidth}
        \centering
        \begin{tabular}{||c c || c c | c c | c c | c c | c ||} 
            \hline
		1T-TMD  & &  $\sigma^{L_y}_{yx}$ & &  $\sigma^{L_z}_{yx}$ & & $|\sigma^L_{yx}|$ & & $|\sigma^S_{yx}|$ & & $E_g (\text{eV})$   \\ [0.5ex] 
		\hline
		\hline
		NiS$_2$ & & $1.02$ & & $-1.78$ & & $2.05$ & & $0.00$ & & $0.54$   \\		
		\hline	        
            PdS$_2$ & & $0.73$ & & $-1.22$ & & $1.42$ & & 0.00 &&$1.14$  \\		
		\hline
		PdSe$_2$ & & $0.85$ & & $-1.65$ & & $1.85$ & & $0.00$ & & $0.52$  \\
		\hline
		PtTe$_2$ & & $0.82$ & & $-1.59$ & & $1.70$ & & $0.00$ & & $0.37$  \\
		\hline
		PtSe$_2$ & & $0.60$ & & $-1.01$ & & $1.18$ & & $0.00$ & & $1.17$  \\
		\hline
	    PtS$_2$ & & $0.51$ & & $-0.77$ & & $0.92$ & & $0.00$ & & $1.72$  \\
		\hline
		ZrS$_2$ & & $0.61$ & & $-0.34$ & & $0.70$ & & $0.00$ & & $1.16$  \\
		\hline
		ZrSe$_2$ & & $0.84$ & & $-0.58$ & & $1.02$ & & $0.00$ & & $0.34$  \\
		\hline
		HfS$_2$ & & $0.32$ & & $-0.08$ & & $0.33$ & & $0.00$ & & $1.27$  \\
		\hline
		HfSe$_2$ & & $0.71$ & & $-0.40$ & & $1.02$ & & $0.00$ & & $0.48$  \\
		\hline
		NiO$^{(*)}_2$ & & $0.51$ & & $-0.76$ & & $0.91$ & & $0.00$ & & $1.25$  \\
		\hline
		PdO$^{(*)}_2$ & &	$0.37$	& &	$-0.46$	& &	$0.59$	& &	$0.00$	& &	$1.39$ \\
		\hline
		PtO$^{(*)}_2$ & &	$0.26$	& &	$-0.27$ & &	$0.37$	& &	$0.00$	& &	$1.70$ \\
		\hline
        \end{tabular}
    \end{minipage}
    \caption{{The heights of the transversal OAM conductivity plateaus for insulating TMD monolayers, which crystallize in the non-centrosymmetric 2H (shown in the left table) and centrosymmetric 1T (shown in the right table) structural phases. The first column displays the stoichiometric formulas for TMD compounds. In the second and third columns, it shows the values of $\sigma^{L_y}_{yx}$ and $\sigma^{L_z}_{yx}$ in units of $e/2\pi$. For both structural phases, $\sigma^{L_x}_{yx} = 0$. The fourth and fifth columns show the values of $|\sigma^S_{yx}|$ and $|\sigma^L_{yx}|$ in unities of $e/2\pi$, where $|\sigma^X_{yx}|=\sqrt{(\sigma^{X_x}_{yx})^2+(\sigma^{X_y}_{yx})^2+(\sigma^{X_z}_{yx})^2}$ for $X=L$ and $S$. The last column presents the energy bandgap for each compound. Details of the numerical calculations of conductivities, along with information on the higher-order topological indexing of each compound, can be found in the supplementary material of Ref. \cite{Costa2023}. 1T-TMDs are HOTIs characterized by $\mathbb{Z}_4=2$ mod $4$ topological indicators. 2H-TMDs are also HOTIs but they are characterized by corner states with a charge $Q^{(3)}_c = 2e/3$ mod $e$. $^{(*)}$ NiO$_2$, PdO$_2$, PtO$_2$ are not TMDs, but they share the same structural, electronic, and topological properties.}}
    \label{table_TMD-Hall}
\end{table*}

\begin{figure*}[t]
    \centering
    \includegraphics[width=0.8\linewidth,clip]{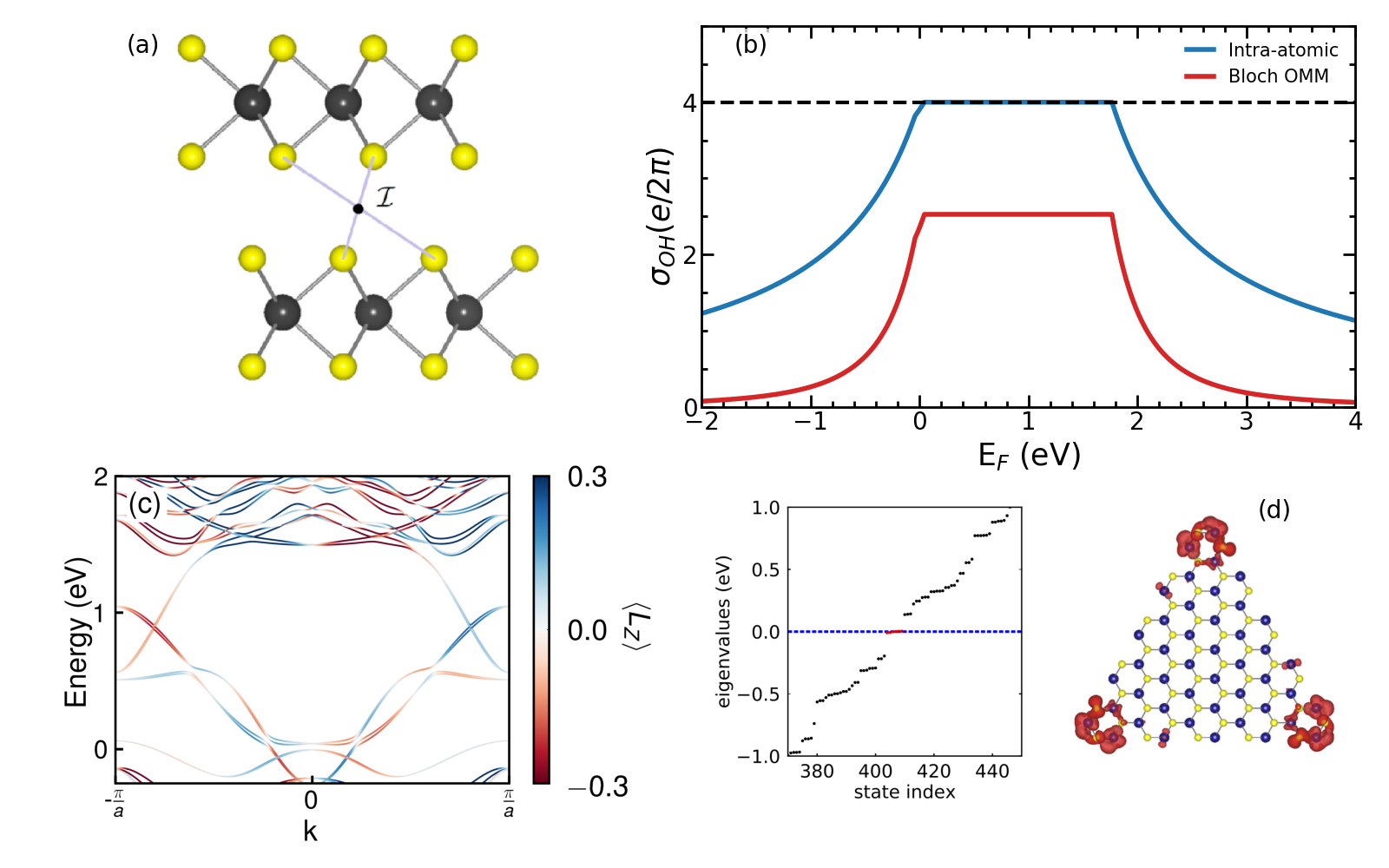}
        \caption{\textbf{Aspects of electronic structure, lattice symmetry, OAM transport, and topology of 2H-TMDs.} (a) Side view of a bilayer of 2H-TMD. $\mathcal{I}$ represents the inversion-symmetry point. (b) The OHC plateau predicted by the low-energy theory for a bilayer of 2H-TMD. The blue curve represents the result obtained using the intra-atomic approximation, while the red curve represents the results derived from the Bloch OMM formulation. The dashed line marks the quantized value of the OHE plateau. (c) The energy spectrum of a MoS$_2$ nanoribbon with zigzag edges reveals OAM-polarized edge states that cross the bulk band and can transport OHC. (d) Energy spectra of triangular nanoflakes of MoS$_2$ with in-gap corner states and the corresponding image of the distribution probability associated with the corner-states wave function. Figures adapted from Refs. \cite{Cysne2021a, Cysne2022, Costa2023}.}
    \label{fig:TMD-top}
\end{figure*}

\subsection*{Orbital v.s. valley transport}
An appealing aspect of orbitronics in 2D materials is the often present multi-valley nature of their low-energy electronic structure. This implies that most of their electronic properties are governed by two inequivalent valleys, ${\bf K}$ and ${\bf K'}$, in the Brillouin Zone (BZ). Such electronic structure is common in many 2D materials with honeycomb structure, such as graphene, hexagonal boron nitride, and 2H-TMDs, among others. The large separation between inequivalent valleys in the BZ suggested that they could be treated as well-defined quantum numbers, allowing them to store information \cite{Xiao2007}. Over the last two decades, this idea has been explored in a field known as valleytronics. However, as reviewed in Ref. \cite{Roche2022}, recent experimental and theoretical studies have revealed persistent inconsistencies with this idea despite all efforts at elucidation. In the standard valleytronic interpretation of typical experiments, an electric field applied to the material produces a transverse valley current via the valley Hall effect. Electrons with distinct valley quantum numbers accumulate at the sample edges. Opposite valley states are characterized by inverted OMM, which can be measured, for instance, using Kerr microscopy. In this view, the accumulation of OMM occurs as a secondary effect consequence of the valley Hall effect. This may seem contradictory because the valley quantum number, which serves as the principal transport quantity, is not directly measured. Instead, the accumulated OMM, typically considered a secondary consequence in valleytronics, can be assessed using Kerr microscopy. Additionally, the definition of valley current ${\bf J}_{v}=-e\left({\bf v}|_{\bf K}-{\bf v}|_{\bf K'}\right)$ depends on a valley filtering procedure that cannot be applied to Hamiltonians with strong inter-valley couplings \cite{Cysne2024a}. Despite the challenges associated with this interpretation \cite{Roche2022}, it has been used to interpret many important experimental results \cite{Lee2016, Wu2013, Wu2019}. 

In Ref. \cite{Bhowal2021}, Bowhal and Vignale argue that the concept of orbital current used in orbitronics could provide a more accurate description than the valley current approach previously utilized in valleytronics. Among other advantages, describing the system in terms of orbital current [using Eqs.(\ref{JLz}) and (\ref{LzBloch})] highlights the OAM—a physical observable probed in experiments—as the primary quantity being transported. Interestingly, orbital and valley current obey distinct symmetry constraints. While the valley Hall effect requires the breaking of time-reversal or spatial inversion symmetry, the OHE does not have such a restriction \cite{Cysne2021a, Cysne2022}. For this reason, the valleytronics community believed that centrosymmetric systems, such as the unbiased bilayer of 2H-TMDs, would not exhibit any transverse current response when a longitudinal electric field is applied. Nevertheless, it was later demonstrated that finite OHC could arise, and the resulting orbital accumulation might be detectable through Kerr rotation experiments \cite{Cysne2021a,Cysne2022}. Even systems with strong inter-valley coupling, such as graphene with Kekulé distortions, can exhibit considerable OHE \cite{Cysne2024a}.

Recently, a study on h-BN/gapped graphene/h-BN moir\'{e} superlattice reported a strong correlation between theoretical calculations of OHC and non-local resistance measurements \cite{Sanchez2024}. Additionally, a recent theoretical study \cite{Sun2024} describing magnetoresistance due to Bloch OMM accumulation in bilayer systems found qualitative agreement with the experiment reported in Ref. \cite{Sala2023}, strengthening the idea of treating it as the primary quantity of transport. Finally, we mention that materials lacking a multivalley structure and SHE, but exhibiting significant OHC, are promising candidates for unambiguously probing orbitronic effects without interference from competing phenomena that might produce similar experimental signals. Phosphorene, a monolayer of black phosphorus, exemplifies such materials and also exhibits highly anisotropic orbital transport properties \cite{Cysne2023b}.

\section{Controlling Orbital Angular Momentum in 2D Materials}
Effects that allow the manipulation of the OAM degree of freedom, such as the orbital Rashba Edelstein effect (OREE), are closely related to the symmetries of solids \cite{Hayami2018, He2020b}. This has been widely studied in the context of spintronics in 2D materials \cite{Offidani2017, Rodriguez-Vega2017} and, as will become evident, offers opportunities to use these systems as platforms for precise OAM control. Additionally, the ability to fabricate heterostructures from different 2D materials can help designing efficient OT devices—a potentiality that has also been explored by the spintronics community of 2D materials \cite{MacNeill2016, Stiehl2019}. This section highlights advances in understanding the phenomena that enable OAM control in 2D materials.

\subsection*{Orbital Rashba Edelstein effect in 2D materials}
In contrast to other orbitronic effects, OREE has been studied both theoretically and experimentally in 2D materials \cite{Xu2021, Son2019, Ye2024, He2020, Li2024b}. Many 2D materials belong to the non-centrosymmetric $D_{\rm 3h}$ point group, where the linear Edelstein effect is forbidden by symmetry [see discussion below Eq. (\ref{OEE2})]. A crucial constraint in 2D materials for allowing OREE is that they can only have one mirror plane perpendicular to the plane of the material. This constraint helps explain why applying strain or cutting the system into a ribbon, thus introducing additional symmetry-breaking elements, can promote the appearance of the OREE. For instance, monolayers of 2H-TMDs, which belong to the $D_{\rm 3h}$ point group, can exhibit a finite current-induced orbital magnetization when their three-fold rotational symmetry is broken through uniaxial strain. This reduces the crystal's point-group symmetry to $C_{2v}$ and enables the OREE, as observed experimentally in Ref. \cite{Son2019} and discussed theoretically in Ref. \cite{Bhowal2020}. Furthermore, fabricating 2D materials in a nanoribbon geometry can reduce their point-group symmetry without the need for a strain field, helping the onset of OREE.\cite{Cysne2023, Cysne2021b}.

Magnetic 2D materials, such as CrI$_3$, exhibit broken time-reversal symmetry, and an equilibrium orbital magnetization may appear. Bilayers of these materials {with interlayer antiferromagnetic configuration} have the interesting properties of breaking time-reversal ($\mathcal{T}\rightarrow \boldsymbol{\times}$) and spatial-inversion ($\mathcal{P}\rightarrow \boldsymbol{\times}$) symmetries individually while preserving their combined operation ($\mathcal{P}.\mathcal{T}\rightarrow \boldsymbol{\checkmark}$). In this case, it is possible to modify the equilibrium orbital magnetization using an electric field without the flow of charge current; that is, a field-induced modification of orbital magnetization can be achieved \cite{Xiao2021}. This implies that even in an insulating state, the OAM can be controlled. This effect resembles the OREE but is governed by distinct symmetry properties \cite{Hayami2018} and a different microscopic mechanism. Along with the OREE, it belongs to the more general class of orbital magnetoelectric effects (see details on nomenclature in Ref. \cite{Osumi2021}). We note that induced orbital magnetization can arise even in higher-order terms in the electric field expansion. This phenomenon corresponds to the non-linear orbital Edelstein effect \cite{Xu2021, Baek2024}. Experiments reporting this effect in 2D materials {are already available} \cite{Ye2024}. Such non-linear phenomena are often studied by the light-matter interaction community, where orbitronics in 2D materials is also under investigation \cite{Mu2021, Li2024, Cysne2024b}.

\subsection*{Torques and OAM in 2D materials}

Orbital torque (OT) has been more extensively investigated in three-dimensional metallic material systems, with only a few studies explicitly addressing torques associated with electronic OAM in 2D materials. In 2D systems, relying solely on the OHE to generate OT is challenging because the OHE induces orbital accumulation at the edges of the plane. However, the orbital Rashba Edelstein effect provides a more effective mechanism by generating orbital magnetization throughout the entire plane. When a 2D magnetic layer is in contact with a material exhibiting OREE, this orbital magnetization can exert torque on the magnetic layer, enabling OT.

Recent theoretical works have begun to explore OAM-related torque phenomena in 2D materials. Calculations by Atencia and co-workers \cite{Atencia2024} suggest the existence of an intrinsic torque on the OAM of Bloch electrons in the presence of an electric field, rooted in interband terms and related to the quantum metric tensor. This torque vanishes in two-band systems with particle-hole symmetric band structures but can be nonzero for tilted Dirac fermions. In Ref. \cite{Canonico2023}, Canonico and co-workers report numerical calculations indicating a spin-orbit torque in the centrosymmetric 1T-TMD PtSe$_2$, arising from the entanglement between spin and the OAM texture of electrons, which is distinct from conventional spin-orbit torques in non-centrosymmetric systems.

Although these studies are not directly related to the OT mechanism introduced in Ref. \cite{Go2020a}, they address phenomena rooted in electronic OAM that could be used for similar purposes. Given the significant impact of spin-torque phenomena on the spintronics community of 2D materials \cite{Liu2020, Tang2021} and the increasing sophistication in fabricating diverse van der Waals heterostructures, it is expected that research into OAM torque phenomena in 2D materials will expand in the coming years. These torque effects may play a crucial role in future applications within these advanced heterostructures \cite{Seyler2018, Zhong2017}.

\section{Effects of Disorder}

Beyond idealized models, real-world applications of orbitronics require {further investigation into the role of disorder, which is prevalent} in the large-scale fabrication of any device. Building on the parallelism between spintronics and orbitronics, critical metrics for addressing the performance of orbitronic devices, such as orbital diffusion lengths and charge-to-orbital conversion {efficiency}, can be altered and even enhanced by disorder. In the following, we separate the orbital relaxation and the extrinsic contributions to the orbital Hall effect and recollect recent experimental and theoretical work on these topics.

\subsection*{Orbital Relaxation and Orbital Dynamics}

{The experimental values of the orbital diffusion length obtained by different methods and for different materials vary considerably, prompting further investigations into} orbital relaxation and the {influence} of disorder. For example, Choi \textit{et al.} \cite{Choi2023}, using MOKE and orbital torque measurements in Ti and Ti/FM systems, obtained orbital diffusion lengths of $74\pm 24$  nm and $50\pm 15$ nm, respectively. {The latter value agrees} with previous estimations {made for} similar heterostructures \cite{Hayashi2023}. In contrast, similar {MOKE} measurements {conducted by} Lyalin \textit{et al.} in Cr reported orbital relaxation lengths of $6.6\pm0.6$ nm \cite{Lyalin2023}. {Similarly,} orbital Hanle {magnetoresistance} measurements of Sala \textit{ et al.} in polycrystalline Mn suggested orbital relaxation lengths of approximately $2$ nm \cite{Sala2023}, which{, surprisingly, }are in close agreement with the estimates of Liu \textit{et al.} in CoPt/Cu/MgO heterostructures \cite{Liu2024}. 

{Although} these experimental data {suggest} that disorder {may limit} orbital conductivity, 
{other studies indicate the opposite.} Kim \textit{et al.} \cite{Kim2023}  
{measured}  orbital relaxation lengths {of approximately} $10$ nm in Cu samples {using orbital torque measurements}. {These findings align with later} estimations for CoFe/Cu/TiO${}_2$ and CoFe/Cu/SiO${}_2$ structures, where the oxygen atoms of the oxides hybridize with copper, favoring orbital transport \cite{Go2021b}. {Similarly,} Seifert \textit{et al.} \cite{Seifert2023} {observed} ballistic orbital propagation {over distances exceeding} $20$ nm, {consistent with earlier} experiments by Hayashi \textit{et al.} \cite{Hayashi2023}.  Furthermore, recent experiments {by} Idrobo \textit{et al.} \cite{Idrobo2024} {examined} orbital relaxation in Ti-based devices and found orbital relaxation lengths of $7.9$ nm, with a strong dependence on sample morphology.
  
The {varying} estimates of orbital diffusion lengths suggest that orbital dynamics {operates} differently from spin dynamics, {which could potentially provide a} method to distinguish between spin and orbital signals.

{Han \textit{et al.} \cite{Han2022} noted that spin and orbital angular momentum operators exhibit distinct anti-comutator relations. Specifically, while $\left\{S_i,S_j\right\}\propto\delta_{ij}$, $\left\{L_i,L_j\right\}$ is proportional to the orbital quadrupole density \cite{Han2024}. These characteristics enable ${\bf L}-{\bf k}$ coupling, but prevent ${\bf S}-{\bf k}$ coupling in centrosymmetric non-magnetic systems. As a result, this can lead to unique features in orbital current diffusion that are not present in spin current diffusion, highlighting qualitative differences between orbital and spin dynamics.}


Sohn \textit{et al.} \cite{Sohn2024} investigated  coupled spin-orbital dynamics in centrosymmetric materials using quantum kinetic theory and device simulations. Their study demonstrated that orbital-momentum locking, arising from the orbital texture, leads to a Dyakonov-Perel-like orbital relaxation, suggesting that frequent momentum scattering disrupts orbital precession, reducing dephasing and thereby increasing relaxation times. Conversely, Ref. \onlinecite{Rang2024} found a different behavior when studying the decay of orbital currents using first-principles {quantum-mechanical} scattering calculations. By injecting orbital currents from an orbitally polarized lead into bulk transition metals, they observed that the orbital current persisted for only a few atomic layers, indicating a relaxation mechanism {different from} the one predicted in Ref. \cite{Sohn2024}.

Recent calculations for a 2D tight-binding model using a quantum Boltzmann equation approach that takes into account different impurity models show that the impurity-{potential} symmetry strongly influences the OAM relaxation{, which} can change from Dyakonov-Perel-like to Elliot-Yafet relaxation if the axial symmetry of the impurity {potential} is broken \cite{Kabanov2024}.

{Despite growing interest in orbitronics, there are currently no direct measurements or systematic studies of the orbital diffusion length in 2D materials. Most of what is known experimentally concerns spin transport in TMDs, where spin diffusion lengths are typically short. However, orbital relaxation mechanisms can differ substantially. For example, in monolayer 2H-TMDs, the OAM near the ${\bf K}$ and ${\bf K}'$ points is strongly constrained by symmetry, which inhibits scattering events that would flip the orbital moment. This symmetry protection suggests the possibility of enhanced orbital diffusion in selected regions of momentum space. While this remains speculative, it motivates further theoretical and experimental work to clarify whether long-range orbital transport can be realized in these systems.}

\subsection*{Disorder effects and Extrinsic contributions to the OHE and OREE}

{Discrepancies in estimates of orbital diffusion lengths also suggest that extrinsic contributions may influence orbital currents and OAM accumulation.}

Since the early days of spintronics, the effects of the disorder have been thoroughly studied and discussed \cite{Sinova2015}. Beyond influencing spin relaxation and decoherence, disorder can also give rise to the SHE through extrinsic mechanisms such as skew scattering and side-jump. These impurity-driven effects can generate spin currents even in materials where intrinsic mechanisms are weak or absent, demonstrating that disorder can play a fundamental role in spin transport phenomena.

In orbitronics, similar extrinsic mechanisms are expected to influence orbital transport, potentially generating OHE. Just as skew scattering and 
side-jump effects can drive {the} SHE, these mechanisms are {also} predicted to {affect} orbital dynamics by deflecting OAM currents in the presence of impurities. {Although} the study of disorder-driven orbital transport is still in its early stages, recent works have begun to explore how extrinsic contributions affect {the} OHE and orbital accumulation. 
{It is noteworthy that these studies deal with effects of disorder on orbital transport  by using either the atom-centered (intra-site) approximation (ACA) or a purely itinerant OAM, which incorporates inter-site contributions.}

{The} seminal paper {of} Bernevig \textit{et al.} \cite{Bernevig2005} discusses the effect of disorder in p-doped semiconductors {within the ACA. They} argued that due to the symmetry of the system {under investigation,} the effects of disorder {appear} as self-energy corrections in their linear response calculations, due to cancellation of the vertex corrections in the current operators. Tanaka \textit{et al.} \cite{Tanaka2008} demonstrated that, for transition metals, the vertex correction considered within the first Born approximation has a negligible impact on the spin and orbital Hall conductivities. Similar trends were obtained in recent work by Pezo \textit{et al.} \cite{Pezo2023}. 

{Recent work by Tang and Bauer \cite{Tang2024} investigate how random defect scattering influences the OHE. They used a quantum Boltzmann equation to examine a two-band ($p_x$-$p_y$) model on a triangular lattice, incorporating the so-called vertex corrections. Their findings reveal that even a weak disorder can significantly affect and may completely suppress the intrinsic orbital Hall current.}

In contrast, for a 2D tight-binding model, a quantum Boltzmann equation approach with third-order perturbation theory shows the appearance of a sizable extrinsic contribution to the OHE. The extrinsic contribution depends on the impurity concentration, in line with the skew-scattering mechanism \cite{Kabanov2024}.

{These results show that the influence of vertex corrections on OHE varies depending on the specific model studied. In principle, the discrepancies can be explained with Dimitrova's argument \cite{Dimitrova2005}, which was originally introduced in the context of the spin Hall effect. This argument is based on general conservation laws that govern the time evolution of the system \cite{Raimondi2011}, which depend on the lattice symmetry and the angular momentum character of the model. For instance, by applying Dimitrova's argument to orbital transport in the $p_x$-$p_y$ model on a triangular lattice, one obtains that the orbital Hall conductivity must vanish, which is consistent with the findings of Ref.\cite{Tang2024}}.

{Meanwhile other studies explored the OHE in two-dimensional massive Dirac fermions, originating from the itinerant OAM. Liu and Culcer \cite{Liu2024b} employed a quantum kinetic approach  to study the effect of short-range Gaussian disorder, taking into account vertex corrections. They discovered that in doped systems, extrinsic Fermi surface contributions can lead to a substantial increase of the OHE, which also aligns with Dimitrova's argument. However, their approach does not capture the dependence of the OHE on impurity concentration and strength.}

Canonico \textit{et al.} \cite{Canonico2024} developed a real-space numerical methodology capable of capturing disorder effects on the OHC non-perturbatively.  {Their results suggest that disorder can enhance the OHE across different transport regimes—from ballistic to diffusive—and eventually suppress the OHE near localization}. 

Further theoretical studies by Veneri \textit{et al.} \cite{Veneri2024}, using a nonperturbative description of disorder in the dilute regime, found that extrinsic contributions to the OHC are dominated by the skew-scattering mechanism. In this regime, the OHC is proportional to the inverse of the impurity concentration, meaning that as impurity concentration decreases, the skew-scattering effect becomes more significant, leading to an increase in OHC. These findings elucidate the origin of extrinsic contributions and emphasize the role of perturbation symmetries in shaping the extrinsic OHC. A subsequent work performed a similar analysis but for centrosymmetric bilayer 2H-TMDs\cite{Faridi2025}.

These findings challenge the prevailing notion that intrinsic mechanisms dominate the OHE, suggesting that extrinsic scattering plays a crucial role, especially for experimentally relevant impurity concentrations.

{Most studies examining the effects of disorder on the orbital Rashba-Edelstein effect adopt the Bloch OMM approach. 
Zhu \textit{et al.} \cite{Zhu2012} derived a general formula for calculating the orbital magnetization in disordered systems, using the Keldysh Green's function formalism. They applied it to calculate the orbital magnetization of a weakly disordered two-dimensional electron gas (2DEG) with Rashba spin-orbit coupling, and found that the orbital magnetization is robust to weak-scattering short-range disorder, which mainly causes a rigid energy shift in the orbital magnetization distribution.} 

{Additionally,} Rou \textit{et al.} \cite{Rou2017} {found through} semiclassical calculations that charge-to-orbital conversion  {has} two extrinsic contributions due to  weak disorder: skew-scattering and side-jump. Although weak disorder can enhance charge-to-orbital conversion, due to its linear dependence on the scattering time, it can be rapidly disturbed in heavy disorder situations.  Further investigation {into} the non-perturbative disordered regime in  2D materials is  {needed}.


\section{Future Directions}

The field of orbitronics in two-dimensional (2D) materials is currently experiencing a stimulating phase, with prospects of development along several lines. However, some challenges need to be addressed to fully realize its potential. One of them is the difficulty in disentangling the contributions of spin and orbital angular momentum (OAM) to magnetic moment accumulations, especially in 2D materials with relatively strong spin-orbit interactions. Extracting these contributions requires the use of more refined experimental techniques, possibly involving optical probes.  There is now direct evidence that OAM currents can be induced and flow through materials even with negligible spin-orbit interaction. Furthermore, these currents can also generate spin currents via spin-orbit coupling (SOC). Therefore, careful interpretation of the experimental results is required to accurately identify and quantify the contributions of OAM to spin-orbitronic processes.

{Another promising line of investigation is exploring the manipulation of hybridization between atoms in 2D materials to enhance the capabilities of orbitronics. For instance, pristine graphene is formed by $sp_2$-hybridized carbon atoms, resulting in primarily $p_z$ states near the Fermi level with no multiorbital character. However, in fully hydrogenated graphene, the carbon atoms undergo $sp_3$ hybridization, which causes the valence band to exhibit a multi-orbital character that can be manipulated through strain \cite{Tokatly2010, Zhang2011}. Furthermore, proximity effects have been successfully employed in spintronics to induce spin-orbit coupling in graphene \cite{Avsar2020}. Similarly, the close proximity of orbitronic materials to conductive materials can lead to new orbital hybridizations \cite{Cysne2018} that may facilitate the transport of OAM currents across heterostructures.
This approach could open up new possibilities for integrating orbitronics into existing 2D material platforms, enabling efficient OAM transport and expanding the scope of potential applications.}

Identifying and engineering 2D materials capable of efficiently transporting OAM currents with minimal damping is critical. Relaxation of OAM poses a significant challenge for the practical implementation of orbitronic devices, as it limits the effective transmission distance of orbital information. Future research should focus on materials exhibiting inherently low OAM relaxation rates or explore strategies to mitigate relaxation through material design or external control. The inherent flexibility of 2D materials makes them particularly well suited for such design optimizations, providing a versatile platform for the development of advanced orbitronic technologies.

{Exploring and optimizing orbital torque in 2D materials is also a promising area for future research. A key step toward demonstrating orbital torque in 2D systems is to achieve orbital accumulation at the interface with a 2D ferromagnet. This could enable torque via orbital-to-spin conversion, or potentially involve direct coupling to orbital magnetization in systems with appreciable unquenched orbital moments—an interesting direction that remains to be explored.} The use of magnetic materials, as sources of OAM currents, is also very encouraging. {The breaking of time-reversal symmetry in the Hamiltonian of magnetic materials results in distinct components in the angular momentum conductivity tensor, which would not be present in a nonmagnetic system. These components can be derived using symmetry arguments \cite{Seemann2015, Wang2021}. They can generate OAM currents with unconventional polarizations, expanding the possibilities for spin-orbitronic applications. A wide range of 2D magnetic materials is currently available, exhibiting electronic properties that range from metallic \cite{Roemer2024, Cardias2025} to insulating \cite{Costa2020, Yu2010,Chang2013}. Exploring new phenomena in these materials will undoubtedly be a key focus within the field of orbitronics.}

The thin and flexible nature of 2D materials makes them ideal for developing compact, low-power devices, where efficient torque generation can enable advances in magnetic switching and data storage technologies.

\section*{Acknowledgements}
The authors thank the Brazilian funding agencies CNPq, CAPES, and FAPERJ for their financial support. TGR
acknowledges financial support from FAPERJ, grant numbers E-26/200.959/2022 and E-26/210.100/2023, FCT - Fundação para a Ciência e Tecnologia, project reference numbers UIDB/04650/2020, 2022.06797.PTDC and 2022.07471.CEECIND/
CP1718/CT0001 (with DOI identifier: 10.54499/2022.07471.CEECIND/CP1718/
CT0001) and 2023.11755.PEX (with DOI identifier https://doi.org/10.54499/2023.11755.PEX) . M.C. acknowledges CNPq (Grant No. 317320/2021-1), FAPERJ/Brazil (Grant No. E26/200.240/
2023) and INCT Materials Informatics for financial support. L.M.C. acknowledges Severo Ochoa Centres of Excellence programme, CEX2021-001214-S and CERCA Programme/
Generalitat de Catalunya. R. B. M acknowledges financial support from INCT of Spintronics and Advanced Magnetic Nanostructures.


%
\end{document}